\documentclass[a4paper,nofootinbib]{revtex4}

\usepackage{amsmath,amssymb}
\usepackage{epsfig}
\usepackage{graphicx}

\newcommand{\be}{\begin{equation}}
\newcommand{\ee}{\end{equation}}
\newcommand{\bi}{\begin{itemize}}
\newcommand{\ei}{\end{itemize}}

\newcommand{\ba}{\begin{eqnarray}}
\newcommand{\ea}{\end{eqnarray}}
\newcommand{\gc}{\gamma_c}
\newcommand{\f}{\frac}

\def\nn{\nonumber \\}

\def\g{\gamma}
\def\d{\partial}
\def\dd{\textrm{d}}
\def\abar{\bar\alpha}

\begin{document}

\title{Asymptotics of QCD traveling waves with fluctuations and running coupling effects}

\author{Guillaume~Beuf}
\email{guillaume.beuf@cea.fr}
\affiliation{Institut de Physique Th\'eorique,
CEA, IPhT, F-91191 Gif-sur-Yvette, France,
CNRS, URA 2306.
}

\begin{abstract}
Extending the Balitsky-Kovchegov (BK) equation independently to running
coupling or to fluctuation effects due to \emph{pomeron loops} is known to
lead in both cases to qualitative changes of the traveling-wave asymptotic solutions. In this paper we study the extension of the forward BK equation, including
both running coupling and fluctuations effects, extending the method
developed for the fixed coupling case
\cite{Brunet:2005bz}. We derive the exact asymptotic behavior in rapidity of the probabilistic distribution of the saturation scale.
\end{abstract}

\maketitle


\section{Introduction}
\label{sec:intro}

In the leading-logarithmic (LL) approximation of high-energy QCD, the evolution with rapidity $Y$ of the dipole-target scattering amplitude which appears in the  dipole factorization of hadronic cross-sections, is given by the Balitsky hierarchy \cite{Balitsky:1995ub}, or by the equivalent JIMWLK equation \cite{Jalilian-Marian:1997jx,Jalilian-Marian:1997gr,Jalilian-Marian:1997dw,Kovner:2000pt,Weigert:2000gi,Iancu:2000hn,Iancu:2001ad,Ferreiro:2001qy}, and in their mean-field approximation, by the Balitsky-Kovchegov (BK) equation \cite{Balitsky:1995ub,Kovchegov:1999yj,Kovchegov:1999ua}. These equations describe the transition from the dilute regime to the dense (saturated) regime of hadronic matter at weak coupling.
The BK equation complements the linear Balitsky-Fadin-Kuraev-Lipatov (BFKL) \cite{Lipatov:1976zz,Kuraev:1977fs,Balitsky:1978ic} equation, whose solutions are exponentially growing, by involving a nonlinear damping term ensuring the taming of the growth, \emph{i.e.} saturation. The Balitsky and JIMWLK equations include correlation effects in the dense regime.

The BK equation has been mapped \cite{Munier:2003vc,Munier:2003sj,Munier:2004xu} onto the Fisher and Kolmogorov-Petrovsky-Piscounov (FKPP) equation \cite{fish,KPP}, which admits universal traveling-wave solutions \cite{bramson, vanS}. This mapping has allowed to rederive some analytical results found in \cite{Mueller:2002zm}, to find various additional results \cite{Munier:2004xu,Marquet:2005qu,Marquet:2005er}, and to give a simple picture of high-energy QCD. In particular \emph{geometric scaling} \cite{Stasto:2000er}, observed in the HERA data, corresponds to a traveling-wave structure of the dipole-target amplitude, and can be understood in term of solutions to the BK equation \cite{Gelis:2006bs}. The geometric scaling is the scaling of the dipole-target amplitude

\be
{\cal N}(Y, k_T^2) \equiv {\cal N}\left(\tau_g = \f{k_T^2}{Q_s^2(Y)}\right)  \, ,
\ee

\noindent $k_T$ being the transverse momentum scale of the dipole, $Q_s(Y)$ the saturation scale, $\tau_g$ the scaling variable.

A key feature of the FKPP equation is that its traveling-wave solutions are driven by the dilute region. Hence these solutions are extremely sensitive to discreteness or disorder effects in the dilute regime which can be added as a noise term to the FKPP equation. The exponential growth in the dilute regime tends indeed to enhance the effect of the fluctuations. The obtained equation is called the stochastic FKPP equation (sFKPP). In particular, if one interprets the FKPP equation as a reaction-diffusion equation for a system of walkers (which we will call \emph{particles}), the particle density can be arbitrary small forward the front of the traveling-wave solutions, even much smaller than the density corresponding to one particle in the whole system, while this very forward region drives the wave solution. The FKPP equation is thus only a mean-field approximation of a reaction-diffusion process, neglecting the relevant fluctuations due to the particle discreteness . These fluctuations can be separated between two types, which have different consequences. First, most of them can be taken into account by cutting off the particle density when it is unphysically small. In \cite{PhysRevE.56.2597}, the modification of the traveling-waves solutions of the FKPP equation in the presence of this cut-off has been derived. The velocity of the wave is reduced w.r.t. the mean field case, and the shape of the front is modified in the dilute region. In the small cut-off limit, the solution is very slowly converging to the mean field solution, typically through inverse of logarithms of the cut-off. Second, in \cite{Brunet:2005bz}, the effect of large fluctuations, not taken into account by the cut-off formalism, has been studied. Those fluctuations, which bring some particles far away forward the wave front, provide a random shift of the front. Hence, the position of the front becomes a random variable.

This problem of sensitivity to the discreteness of the interacting particles has been raised in the QCD context in \cite{Salam:1995zd,Salam:1995uy,Salam:1996nb,Mueller:1996te} for the simulations of onium scattering with unitarity corrections, and discussed in \cite{Iancu:2003zr,Iancu:2004iy} where they are associated with pomeron loops. This has led to conjecture that high-energy scattering processes in QCD may fall in the same universality class as a reaction-diffusion process \cite{Iancu:2004es}, extending the mapping from BK to FKPP beyond the mean field approximation. The generalization of the Balitsky and JIMWLK equations including pomeron loops have been investigated\footnote{The aim of these studies is not only to include fluctuations, but also to build a formalism for saturation with a symmetric treatment of projectile and target. } through various formalisms \cite{Braun:2000bi,Braun:2005hx,Iancu:2004iy,Iancu:2005nj,Mueller:2005ut,Levin:2005au,Kovner:2005nq,Kovner:2005en,Balitsky:2005we,Hatta:2005rn}.  For example, the QCD evolution equation with pomeron loops has been formulated in \cite{Iancu:2004iy,Iancu:2005nj} as a Langevin equation. If one neglects the non-locality of the dipole-dipole interactions and performs a coarse-graining approximation, the evolution equation with pomeron loops can be rewritten as a simpler Langevin equation\footnote{The approximations leading from the full equation with pomeron loops to \eqref{sBK} are not well understood. Moreover, some doubt has been cast on the validity of the approximation \eqref{sBK} for QCD \cite{Iancu:2005nj,Kozlov:2007xc}.   However, \eqref{sBK} is a useful toy model to investigate the fluctuation effects in QCD, and we will keep this approximation in our work.} \cite{Iancu:2004iy}, namely

\be \label{sBK}
\d_Y {\cal N}(L,Y) = \abar \left[ \chi(-\d_L) {\cal N}(L,Y)-{\cal N}^2(L,Y)+\sqrt{\kappa \alpha_s^2 \ {\cal N}(L,Y)} \ \eta(L,Y) \right] \; ,
\ee

\noindent where ${\cal N}$ is the forward dipole-target scattering amplitude, $L$ is related to the dipole momentum scale $k_T$ and some reference scale $Q_0$ by $L\equiv \log (k_T^2 / Q_0^2)$, $\kappa$ is a coarse-graining factor taking into account the approximation, and $\abar \equiv N_c \alpha_s / \pi$. $\chi(-\d_L)$ is the LL BFKL kernel restricted to zero momentum transfer and zero conformal spin, with the eigenvalues $\chi(\g)=2\Psi(1)-\Psi(\g)-\Psi(1\!-\!\g)$. $\eta$ is a Gaussian white noise with the correlation

\be
\langle \eta(L,Y) \, \eta(L',Y') \rangle = \f{4}{\abar} \ \delta(L-L') \delta(Y-Y') \; ,
\ee

\noindent and the equation \eqref{sBK} should be understood with the Ito prescription\footnote{See for example \cite{vanKampen} or \cite{Gardiner} for a review.}.  Our goal will be to study this equation with running coupling instead of fixed.

The main prediction of the fluctuations studies at fixed coupling is the existence of \emph{diffusive scaling} at high rapidity. The fluctuations indeed seems to randomly shift the wave front. Hence, geometric scaling holds event by event but is no more valid in average, and is replaced by diffusive scaling, \emph{i.e.}

\be\label{diffusivescFC}
{\cal N}(Y, k_T^2) \equiv {\cal N}\left(\tau_d = \f{\log \left( k_T^2 / Q_s^2(Y)\right)}{\sqrt{D Y}}\right)  \, ,
\ee
\noindent $D Y$ being the variance of the random saturation scale. Moreover, the averaging reduces the slope of the amplitude. The diffusive scaling has been conjectured in \cite{Iancu:2004es}, and derived in the small noise limit in \cite{Marquet:2006xm}, using the results of \cite{Brunet:2005bz} applied to the equation \eqref{sBK}. It has been also derived in the strong noise limit \cite{Marquet:2005ak}. Numerical simulations indicates that this diffusive scaling holds in the case of a finite noise term \cite{Soyez:2005ha,Iancu:2006jw}. The diffusion coefficient $D$ is a growing function of the  strength of the noise.

Almost all the above-mentioned studies have been performed at LL accuracy. But the extension of the saturation equations beyond LL level is one hot topic. A part of the Next to Leading Logarithm (NLL) corrections to the BK equation is already known \cite{Balitsky:2006wa,Kovchegov:2006vj,Kovchegov:2006wf,Albacete:2007yr}. In general, some of the NLL corrections contribute to the running coupling and some other to the kernel and to the nonlinear terms. If one knows them, one can resum higher order contributions to the running of the coupling, and obtain an equation containing the running coupling at the relevant scale and NLL kernel and nonlinear terms. The precise shape of the nonlinear terms is not relevant at high rapidity in the dilute regime, due to universal properties of the traveling waves solutions. It has also been shown \cite{Beuf:2007cw} that the NLL corrections to the kernel are not relevant for the transition to the saturation regime at high enough rapidity in the running coupling case. By contrast, the replacement of the fixed coupling by the running coupling has a significant impact on the asymptotic solutions \cite{Mueller:2002zm,Triantafyllopoulos:2002nz,Munier:2003sj}. In particular, one has at large rapidity $\log Q_s^2(Y) \propto \sqrt{Y}$, instead of $\log Q_s^2(Y) \propto Y$ in the fixed coupling case.

As the fluctuations and the running coupling seem to be both relevant for the saturation at high energy, they have \emph{a priori} to be both taken into account\footnote{A first attempt in that direction can be found in \cite{Mueller:2004se}.}. The purpose of the present work is to study the solution of a model extending the BK equation, with both running coupling and fluctuations effects.

We explain in section \ref{sec:formul} how we implement the running coupling in \eqref{sBK} to build our model. Then, we apply the method developed in \cite{Brunet:2005bz} to our model, implementing the cut-off in section \ref{sec:cutoff} and the large fluctuations in section \ref{sec:largef}. Our final result is presented in section \ref{sec:results}. In concluding section \ref{sec:conclu}, we compare our result to the one in the fixed coupling case, and discuss the relevance of our model.


\section{Formulation of the model}\label{sec:formul}

Let us take into account the running coupling effects in the equation \eqref{sBK}. As mentioned in the introduction, the full NLL corrections to the BK equation are being calculated, but not yet available. However they are not all contributing to the asymptotic behavior of interest in the present paper \cite{Beuf:2007cw}. Even the precise choice of the momentum scale for the running coupling will have no effect on our results\footnote{\emph{A priori}, the scales present in the problem are the size the parent and daughter dipoles. The first NLL parts calculated for the BK equation give hints to determine the running coupling scale \cite{Balitsky:2006wa,Kovchegov:2006vj}, except for the fluctuation term, which does not come from the typical fan diagram structure of the BK equation. However, only a rough estimate of the running coupling scale is needed, because the result of the present paper is quite robust with respect to a change of the running coupling scale, especially for the $\alpha_s^2$ coupling present in the fluctuation term. In order to check the independence of our result from this choice, the following analysis is done in the appendix \ref{sec:PDM} with the choice of the parent dipole momentum, \emph{i.e.} the Fourier conjugate of the parent dipole size, as the running coupling scale.}.
In the following, we take the running coupling at the saturation scale everywhere in the equation \eqref{sBK}, namely

\be
\abar^{-1} \equiv b \ \rho_s(Y) \quad \textrm{where} \quad b=\f{11 N_c - 2 N_f}{12 N_c} \quad \textrm{and} \quad \rho_s(Y) \equiv \log (Q_s^2(Y) / \Lambda_{QCD}^2) \; . \label{RCsatscale}
\ee

With the assumption \eqref{RCsatscale}, the evolution equation \eqref{sBK} becomes

\ba \label{sBKr1}
b \ \rho_s(Y) \ \d_Y {\cal N}(L,Y) &=&   \chi(-\d_L) {\cal N}(L,Y)-{\cal N}^2(L,Y)+\sqrt{\f{\kappa \pi^2 \  {\cal N}(L,Y)}{N_c^2 b^2 \ \rho_s^2(Y)}} \ \eta(L,Y)  \; , \\
\textrm{with}  & &  \langle \eta(L,Y) \, \eta(L',Y') \rangle = 4 b \  \rho_s(Y) \ \delta(L \! - \! L') \delta(Y \! - \! Y')\nonumber \; ,
\ea

\noindent where $L$ is now defined by $L\equiv \log (k_T^2 / \Lambda_{QCD}^2)$.
As $\rho_s(Y)$ is a priori unknown, one start with the assumption that \eqref{sBKr1} has indeed a traveling front solution, $\rho_s(Y)$ being the position of the front in $L$ variable, and then check for consistency. Thus $\rho_s(Y)$ and ${\cal N}(L,Y)$ can be both determined by \eqref{sBKr1}.

Let us start our analysis by considering the Ansatz $\rho_s(Y) \equiv v(Y^\beta) \ Y^\beta$ where $v(t)$ has a finite and non-zero limit for $t \rightarrow \infty$, $t\equiv Y^\beta$ being the \emph{effective time} of the wave propagation. Writing the equation \eqref{sBKr1} in the $t$ variable, one remarks that the only consistent value of $\beta$ is $1/2$ (as expected from the previous studies of BK equations with running coupling \cite{Mueller:2002zm,Munier:2003sj}). Then, the evolution equation writes

\ba \label{sBKr}
\f{b \ v(t)}{2} \ \d_t N(L,t) &=&   \chi(-\d_L) N(L,t)-N^2(L,t)+\sqrt{\f{\bar\kappa \  N(L,t)}{v(t) \ t^2}} \ \nu(L,t)  \; , \\
\textrm{with}  & &  \langle \nu(L,t) \, \nu(L',t') \rangle = \delta(L \! - \! L') \delta(t \! - \! t') \; ,\nonumber
\ea

\noindent where $t\equiv \sqrt{Y}$, $N(L,t) \equiv {\cal N}(L,Y=t^2)$, $\nu(L,t) \equiv \eta(L,t^2) \ / \sqrt{2b \ v(t)}$, and $\bar\kappa \equiv 2 \kappa \pi^2 / N_c^2 b\ $.


\section{Derivation of the solution}

\subsection{Cut-off contribution} \label{sec:cutoff}

The fluctuation term in \eqref{sBKr} is relevant only at very small density, namely for $N(L,t) \lesssim \bar\kappa / v(t) \ t^2$. Hence, the shape of the wave front should not be modified at larger densities. However, in the standard BK equation case, the selection of the wave velocity happens forward the front, where the density is very small. One expects that the fluctuations modify the forward part of the front, and thus the velocity of the wave front.
Following \cite{PhysRevE.56.2597}, we replace the fluctuation term in \eqref{sBKr} by a cut-off, which corresponds to a deterministic approximation for the bulk of the fluctuations (see Fig. 1). When the density is small, a fluctuation may indeed set the density back to zero where it stays since zero is a fixed point.

\begin{figure}
\begin{center}
\includegraphics[width=10cm]{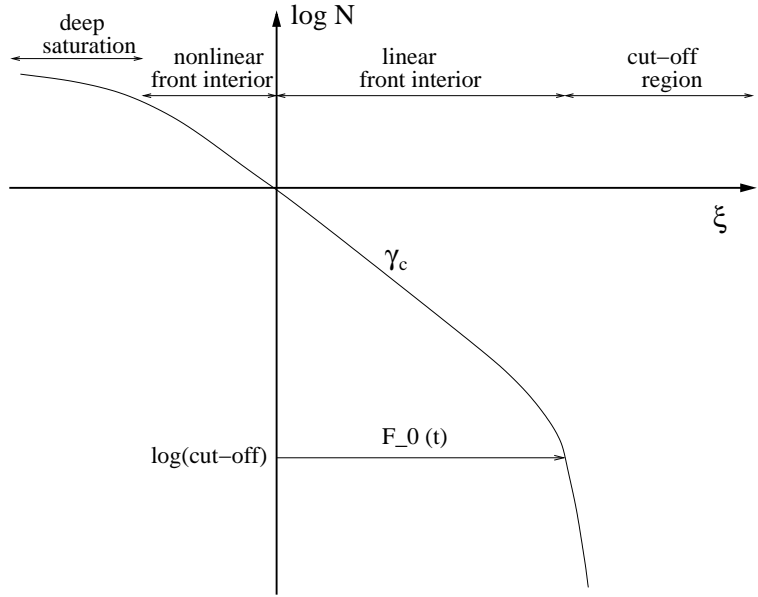}
\label{fig:cutoff}
\caption{Shape of the solution in the presence of the cut-off, in the comoving frame of the wave front.}
\end{center}
\end{figure}

We first study the \emph{linear} part of the \emph{front interior} region (see Fig. 1), defined by
\be
\f{\bar\kappa}{ v(t) \ t^2} < N(L,t) \ll 1 \, ,
\ee

\noindent where we can neglect both the nonlinear and fluctuation terms. In this region

\be
\f{b \ v(t)}{2} \ \d_t N(L,t) \simeq   \chi(-\d_L) N(L,t) \, .
\ee

We now switch to the comoving frame of the front, replacing $L$ by $\xi \equiv L - v(t) t + \xi_0$ and $\bar N(\xi, t)\equiv N(L,t)$, $\xi_0$ being some parameter. Then

\be \label{sBKr11} \f{b \ v(t)}{2}[\d_t \bar{N}(\xi, t) -(v(t)+t \dot{v}(t))\d_\xi \bar{N}(\xi,t)]=\chi(-\d_\xi) \bar{N}(\xi,t) \, .
\ee

\noindent The free parameter $\xi_0$ is fixed by imposing that $\xi=0$ corresponds to the separation between the regions where the nonlinearity is relevant or not (see Fig. 1). Then, the linear part of the front interior is defined by $0 < \xi < F_0(t)$,  $F_0(t)$ being the position of the cut-off in the comoving frame.
In \cite{PhysRevE.56.2597}, the velocity has been found decreased by a term depending on the cut-off through logarithm. In our case, the cut-off is decreasing like $t^{-2}$ because of the running of the coupling. Hence, we can expect that the velocity is decreasing through powers of logarithm of $t$. We thus drop the time derivative term in \eqref{sBKr11}, which is decreasing like $t^{-1}$. We obtain the differential equation

\be \label{sBKr12} -\f{b \ v(t)}{2}(v(t)+t \dot{v}(t)) \ \d_\xi \bar{N}(\xi,t)=\chi(-\d_\xi) \bar{N}(\xi,t) \ee

\noindent whose solutions are

\be \label{soleq}
\bar{N}(\xi,t)= \sum_i C_i \ e^{-\gamma_i(t) \xi} \, ,
\ee

\noindent where the $\gamma_i(t)$ are the solutions of the dispersion relation

\be \label{disp} \f{b}{2} v(t) (v(t)+t \dot{v}(t)) = \f{\chi(\gamma(t))}{\gamma(t)} \, . \ee

As the fluctuation term is vanishing for $t\rightarrow \infty$, $\gamma(t)$ and $v(t)$ should converge to the critical values $\gamma_c$ and $v_c$ of the BK equation with running coupling \cite{Munier:2003sj}, namely

\be \label{param} \chi(\gc) = \gc \ \chi'(\gc)  \ , \quad {\textrm{and}} \quad v_c \equiv \sqrt{\f{2 \chi(\gc)}{b \gc}} \; \; . \ee

Following the method of \cite{PhysRevE.56.2597}, we expand the dispersion relation \eqref{disp} around these critical values, and find

\be \label{edisp} -b v_c \delta v(t) -\f{b v_c}{2} \  t \delta \dot{v}(t) + {\cal{O}}((\delta v(t))^2) = \f{\chi''(\gc)}{2 \gc} \  (\delta \g(t))^2 + {\cal{O}}((\delta \g(t))^3) \, , \ee

\noindent where $\delta \g (t) \equiv \g (t) - \gc$ and $\delta v(t) \equiv v_c - v(t)$ ($\delta v>0$, since the cut-off is expected to reduce the velocity). We have supposed that $\delta v(t) \propto (\log t)^{-\zeta}$ with some positive $\zeta$, then $t \ \delta \dot{v}(t) \propto (\log t)^{-\zeta-1}$ is a subleading term at very large $t$ in the expansion \eqref{edisp}. Hence

\be \delta \g (t) = \pm i \pi \sqrt{\f{\delta v(t)}{\lambda}} \ , \quad {\textrm{where}} \quad \lambda \equiv \f{\pi^2 \chi''(\gc)}{2 b \gc v_c} \, . \ee

We can rewrite at large $t$ the solutions \eqref{soleq} as

\be \label{soleq1} \bar{N}(\xi,t)=e^{-\gc \xi} \left[ C \sin\left(\pi \sqrt{\f{\delta v(t)}{\lambda}} \xi\right) + D \cos\left(\pi \sqrt{\f{\delta v(t)}{\lambda}} \xi\right)\right] \, . \ee

The shape of the front at small $\xi$, away from the cut-off, has to be the same as the one without fluctuations

\be \label{fishapemf} \bar{N}(\xi,t) \sim A \  \xi e^{-\gc \xi} \, . \ee

One finds the constants $C$ and $D$ by matching \eqref{soleq1} with \eqref{fishapemf}, and writes

\be \label{fishapeco} \bar{N}(\xi,t) = \f{A}{\pi } \ \sqrt{\f{\lambda}{\delta v(t)}} \  \sin\left(\pi \sqrt{\f{\delta v(t)}{\lambda}} \xi\right) e^{-\gc \xi } \quad \textrm{for} \quad 0 \leq \xi \leq F_0(t) \quad \textrm{and $\ t \gg 1$ .}  \ee

At the cut-off, one has

\be \label{contco} \bar{N}(F_0(t),t) = \f{A}{\pi } \ \sqrt{\f{\lambda}{\delta v(t)}} \  \sin\left(\pi \sqrt{\f{\delta v(t)}{\lambda}} F_0(t)\right) e^{-\gc F_0(t) } = \f{\bar\kappa}{v(t) \ t^2} \, . \ee

As $F_0(t)$ and $\delta v(t)$ are both unknown, we have also to ensure the continuity of the derivative of the front at $F_0(t)$ in order to determine them both. The idea of the cut-off method is to prevent the front from having a too large tail in the dilute region. So let us impose at $F_0(t)$ a stronger slope $\bar\g > \gc$, with for example $\bar\g-\gc \simeq 1$. This gives a second equation

\be \label{derco}
\left[-\f{\gc A}{\pi} \  \sqrt{\f{\lambda}{\delta v(t)}} \  \sin\left(\pi \sqrt{\f{\delta v(t)}{\lambda}} F_0(t)\right) + A \cos\left(\pi \sqrt{\f{\delta v(t)}{\lambda}} F_0(t)\right) \right] \  e^{-\gc F_0(t) } = -\f{\bar{\g} \bar\kappa}{v(t) \ t^2}
\ee

The  equations \eqref{contco} and \eqref{derco} are similar to the ones in \cite{PhysRevE.56.2597}, and gives the following result
\ba
v(t) &=& v_c - \f{\lambda}{F_0^2(t)} = v_c - \f{\pi^2 \chi''(\gc)}{2 b \gc v_c F_0^2(t)}\label{vco1} \\
\bar{N}(\xi,Y)&=&\f{A F_0(t)}{\pi} \  \sin\left(\f{\pi \xi}{F_0(t)}\right) e^{-\gc \xi } \quad \textrm{for} \quad 0 \leq \xi \leq F_0(t) \quad \textrm{and $t$ large} \label{frontco1}\\
F_0(t) &=& \f{1}{\gc} \log \left( \f{A v(t) t^2}{(\bar{\g}-\gc) \bar\kappa} \right) = \f{2}{\gc} \log t + \f{1}{\gc} \log \left( \f{A v_c }{(\bar{\g}-\gc) \bar\kappa} \right) + {\cal O}(\delta v(t)) \nonumber
\ea
These results have only a weak dependence in the free parameter $\bar{\g}$ and in the coarse-graining factor $\kappa$. In fact, as we have already truncated asymptotic series of $\log t$ in \eqref{edisp}, only the first term is really significant. Thus, our result is 
\ba
F_0(t) &=& \f{2}{\gc} \log t \quad \textrm{for the width of the front,}\label{F0}\\
v(t) &=&  v_c - \f{\pi^2 \gc \chi''(\gc)}{8 b  v_c \log^2 t} \quad \textrm{for the velocity of the front,}\label{vco} \\
\bar{N}(\xi,Y)&=&\f{2 A \log t}{\pi \gc} \  \sin\left(\f{\pi \gc \xi}{2 \log t}\right) e^{-\gc \xi } \quad \textrm{for the shape of the front, for} \; 0 \leq \xi \leq \f{2}{\gc} \log t \; \textrm{and $t$ large}\, . \label{frontco} 
\ea


\subsection{Contribution from large fluctuations} \label{sec:largef}

As explained in \cite{Brunet:2005bz}, the cut-off method allows to take into account the frequent but small fluctuations, which can eventually set $N$ to zero. But the large fluctuations, which can give a non-zero value to $N$ far away forward the front, are too rare to be modeled in a deterministic way. However they can have an important effect. The physical picture is the following (see Fig. 2): the traveling-wave is essentially given by the cut-off contribution, but sometimes, a fluctuation brings a few particles forward the tip of the front, setting there $N$ of the same order as the cut-off. Then, this seed grows, forming a new forward-wave front, and also a backward-wave front. Finally, the old wave front meets the backward-wave front, and only the forward front created by the fluctuation remains. The long-time effect of a large fluctuation is thus to shift the wave front a bit forward.

\begin{figure}\label{fig:fluct}
\begin{center}
\begin{tabular}{ccc}
\includegraphics[width=7cm]{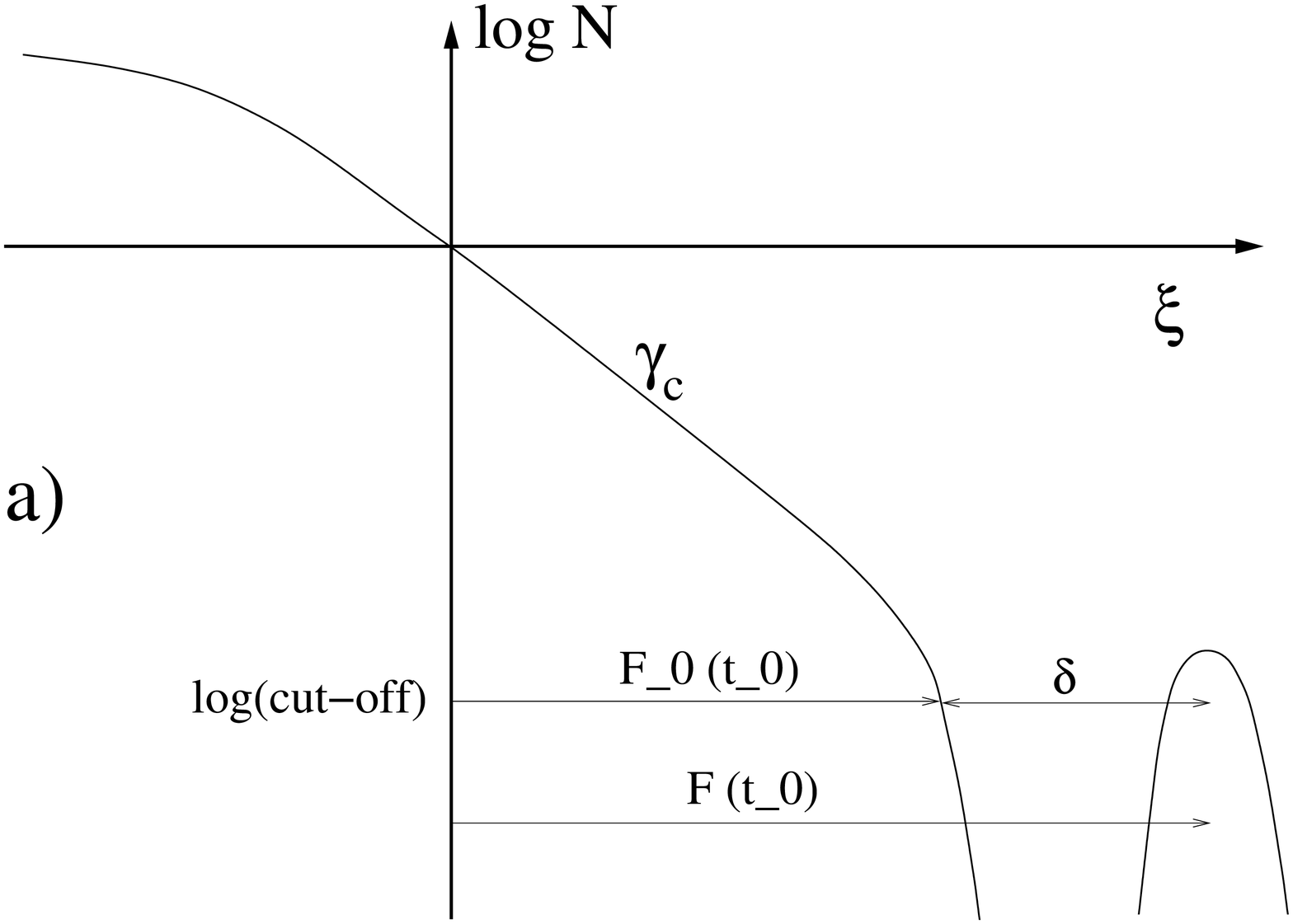} & & \includegraphics[width=7cm]{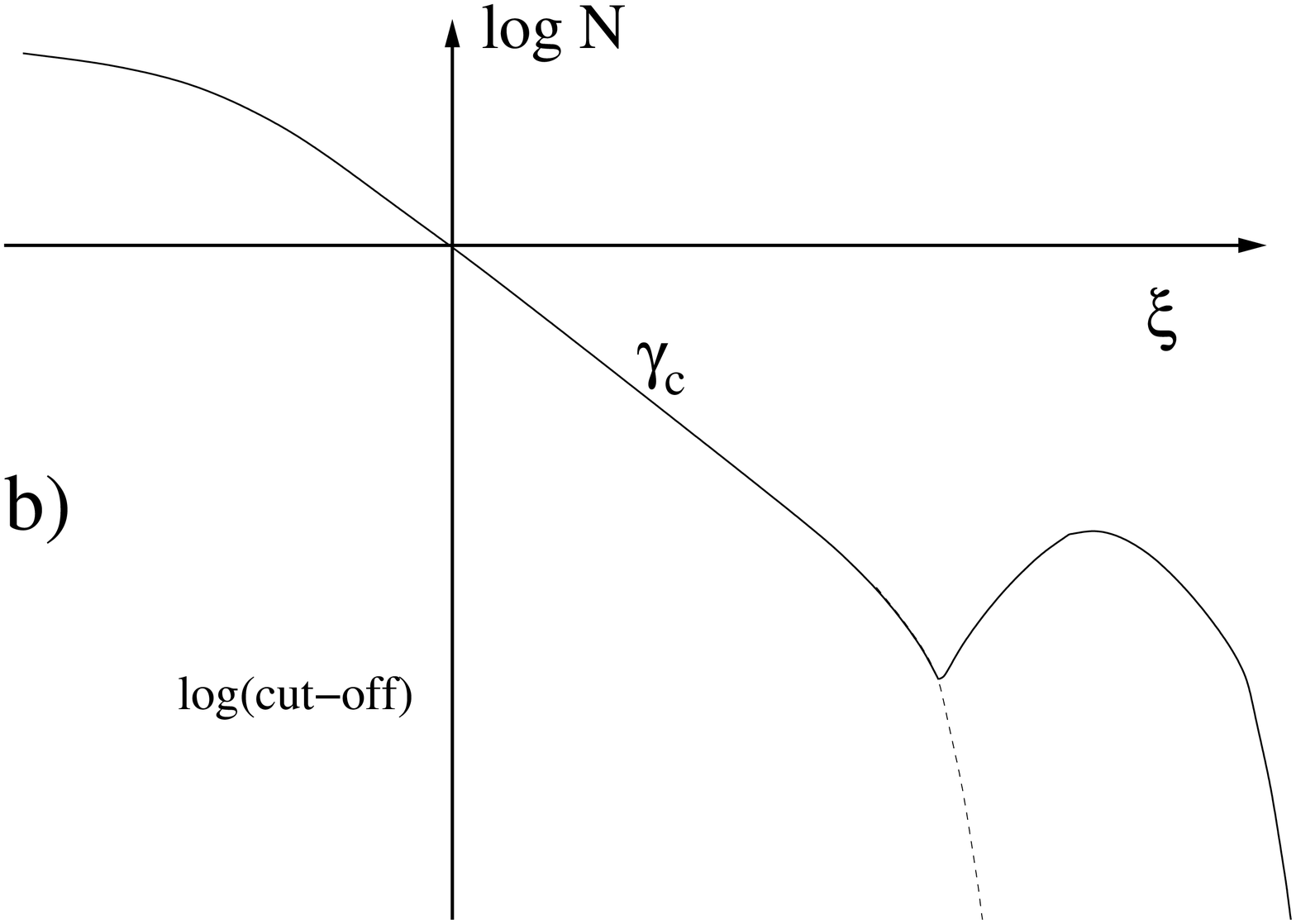}\\
& & \\
\includegraphics[width=7cm]{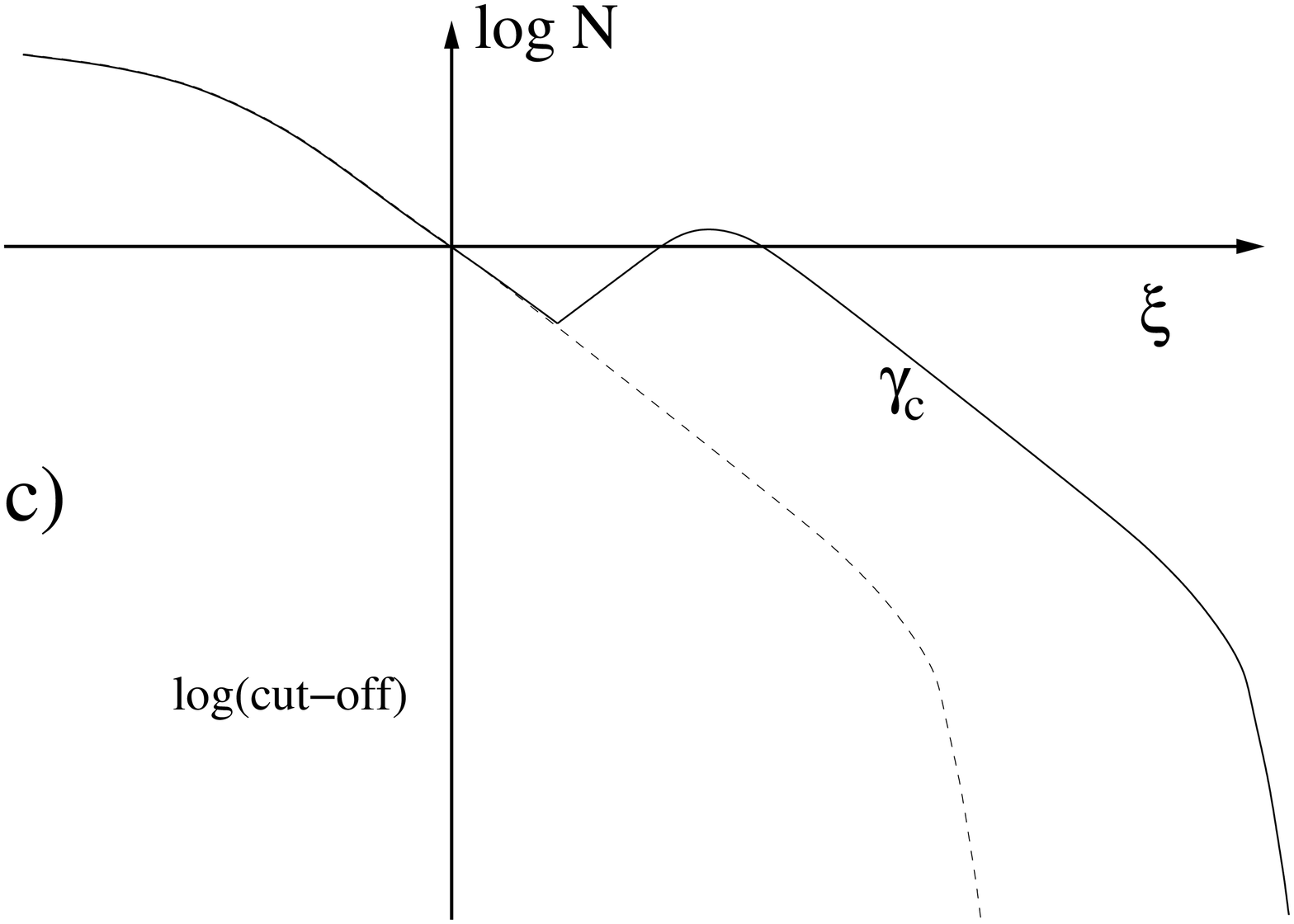} & & \includegraphics[width=7cm]{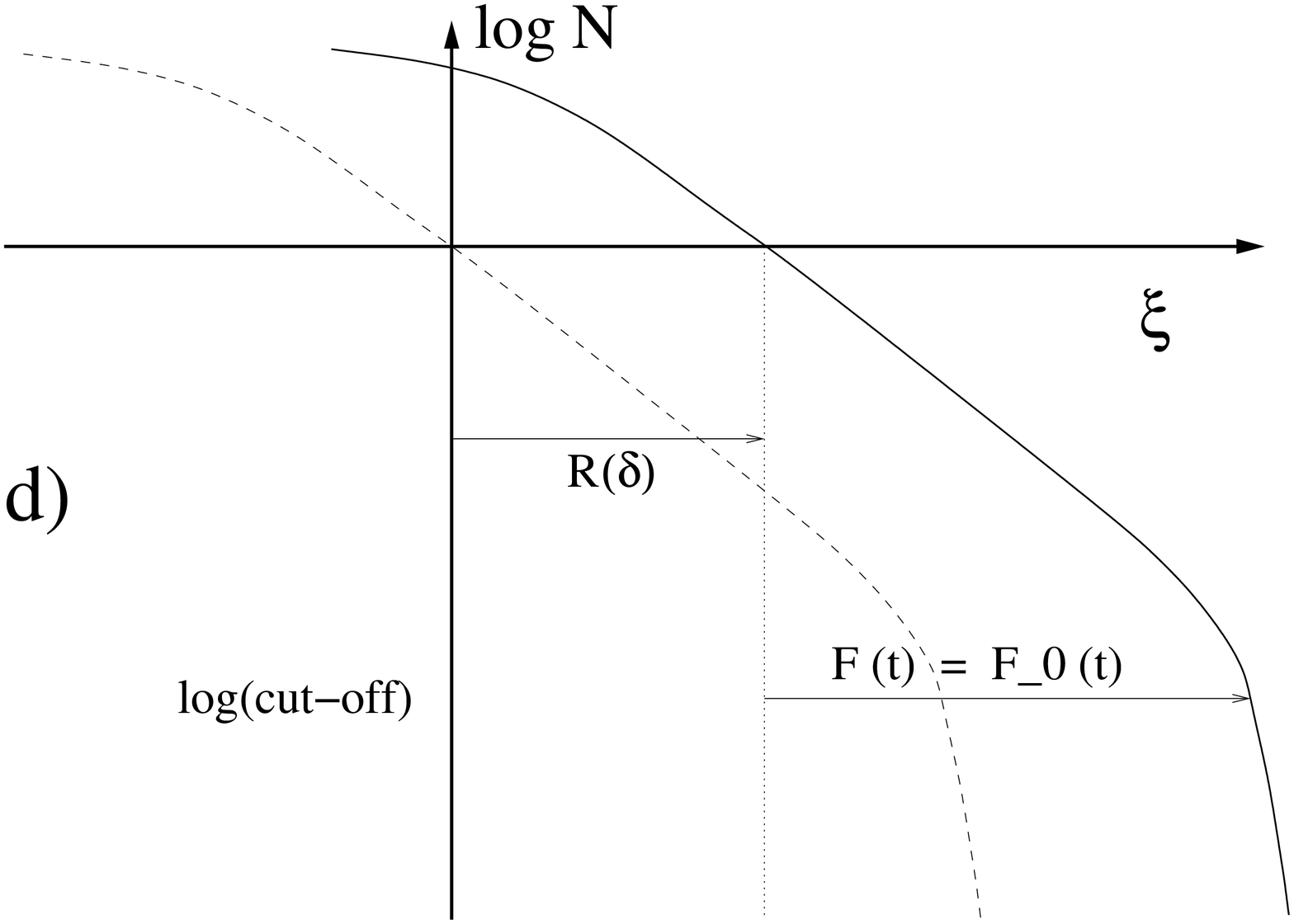}
\end{tabular}
\caption{Relaxation of the front after the creation of a seed by a large fluctuation. a) shows the initial condition with the cut-off front and the seed, b) the growth of the seed and c) the merging between the fronts. d) represents the final state of the relaxation, with a front similar the initial one (in dashed line), but translated by $R(\delta)$. Note that the solution is drawn in the comoving frame with the velocity determined with the cut-off formalism.}
\end{center}
\end{figure}

The first assumption in \cite{Brunet:2005bz} is that such a fluctuation has a probability to happen during $\dd t$ at a distance between $\delta$ and $\delta +\dd \delta$ forward the front

\be \label{probdelta}
{\cal P}(\delta) \ \dd \delta \ \dd t = C_1 e^{-\gc \delta} \ \dd \delta \ \dd t \, ,
\ee

\noindent with $C_1$ being some unknown normalization constant.
This probability law depends on the shape of the front because the noise term in the sFKPP equation depends on the particle density, as in equation \eqref{sBKr}. We keep the assumption \eqref{probdelta} in the running coupling case\footnote{The assumption \eqref{probdelta} has been verified numerically in the sFKPP case, and the physical arguments supporting it hold for the equation \eqref{sBKr}.}.

Following \cite{Brunet:2005bz}, one has to study the effect of one single fluctuation, \emph{i.e.} the relaxation of the front, starting from an initial condition at $t=t_0$ corresponding to the cut-off traveling-wave solution plus a seed at a distance $\delta$ (see the first picture in Fig. 2). $t_0$ is taken very large, so that $F_0(t_0)$ is large. The effective width of the front interior $F(t)$ is larger than the one for the cut-off solution $F_0(t)$, starting at $F(t_0)=F_0(t_0)+\delta$.
In the comoving frame (with the velocity \eqref{vco}), this relaxation is due to diffusion on the interval $[0,F(t_0)]$. Hence, the typical relaxation time is of order $(F_0(t_0))^2$. For the linear front interior up to the fluctuation, we take a similar Ansatz as in \cite{Brunet:2005bz}, namely

\be \label{Ansatz}
N(L,t) = F(t) \ G\left(\f{L\! - \! v_e(t)t +\xi_0}{F(t)},\f{t-t_0}{(F_0(t_0))^2} \right) e^{-\gc (L-v(t)t +\xi_0)} \, ,
\ee

\noindent where $v_e(t)$ is an effective velocity of the front, such that

\be
v_e(t_0)=v(t_0) \quad \textrm{and} \quad R(\delta) \equiv \lim_{t\rightarrow \infty} [v_e(t)t\! - \! v(t)t] \, ;
\ee

\noindent $R(\delta)$ being the shift of the wave front induced by the fluctuation. As $v_e(t)t\! - \! v(t)t$ evolves only during the relaxation, and $R(\delta)$ is of order $1$, as $\delta$, one has

\be \label{vderiv}
\d_t \left[v_e(t)t\! - \! v(t)t\right] \propto (F_0(t_0))^{-2} \, .
\ee

As $\gc \simeq 0.6275$,  $F(t)-F_0(t)$ starts of order $1$, according to \eqref{probdelta}, and relaxes to $0$. Hence, one infers

\be \label{Fderiv}
 \d_t \ [F(t)-F_0(t)] \propto (F_0(t_0))^{-2} \quad \textrm{and then} \quad  \d_t \ F(t)  \propto (F_0(t_0))^{-2} \, .
\ee

Inserting the Ansatz \eqref{Ansatz} in

\be
\f{b \ v_e(t)}{2} \ \d_t N(L,t) =   \chi(-\d_L) N(L,t) \label{relaxeq}
\ee

\noindent with the assumptions (\ref{vderiv},\ref{Fderiv}), one gets

\be \label{eqtrans}
b v_c \ \d_\tau G(y,\tau) - \pi^2 \chi''(\gc) \ G(y,\tau) = \chi''(\gc) \ \d_y^2 G(y,\tau) \,  ,
\ee

\noindent where

\ba
y &\equiv& \f{L\! - \! v_e(t)t +\xi_0}{F(t)} = \f{\xi\! - \! (v_e(t)t\! - \!v(t)t)}{F(t)}\nn
\tau &\equiv& \f{t-t_0}{(F_0(t_0))^2}  \, .
\ea

We take as boundary conditions
\be \label{bcy}
G(0,\tau)=G(1,\tau)=0 \, ,
\ee

\noindent because the solution has to match \eqref{fishapemf} at small $y$ in order to take the nonlinear term into account, and $y=1$ corresponds to the position of the cut-off for the wave front generated by the fluctuation. The value of the cut-off gives a boundary condition for $G$ too small to be relevant within the accuracy of our calculations.

The equation \eqref{eqtrans}, with \eqref{bcy}, is the same as for the FKPP case and for the BK equation with fixed coupling, after a rescaling of $\tau$. Hence, as in these cases, the first Fourier mode of G is constant, and the others decay exponentially. Calculating this first Fourier mode for the chosen initial condition, and matching it with the cut-off solution \eqref{frontco} translated by $R(\delta)$, one finds

\be \label{shiftfront}
R(\delta)=\f{1}{\gc} \ \log \left( 1+ \f{C_2 e^{\gc \delta} }{(F_0(t_0))^3}\right) \, ,
\ee

\noindent where $C_2$ is some unknown constant of order $1$.

We turn now to the problem of the long time behavior of the solution, including several large fluctuations.
\eqref{shiftfront} tells us that a fluctuation is relevant if $e^{\gc \delta}$ is at least of order $(F_0(t_0))^3$. Hence, using \eqref{probdelta}, one can conclude that the typical time between two relevant fluctuations is of order $(F_0(t))^3$. For $t$ large enough, this time is much larger than the relaxation time of a large fluctuation, which is of order $(F_0(t))^2$. Hence we can admit that the wave front has enough time to completely relax between two large fluctuations.

The calculation of the time derivatives of the cumulants of the position of the front done in \cite{Brunet:2005bz} holds, thanks to this property, to \eqref{probdelta}, and to \eqref{shiftfront}. It gives

\ba \label{derivcumnn}
\d_t \langle \rho_s(t^2) \rangle &=& v_c - \f{\pi^2 \chi''(\gc)}{2 b \gc v_c F_0^2(t)} +\f{C_1 C_2}{\gc^2} \ \f{3 \log F_0(t)}{(F_0(t))^3} \nn
\d_t \langle \rho_s^n(t^2) \rangle_c &=& \f{C_1 C_2 \ n! \zeta(n) }{\gc^{n+1} \ (F_0(t))^3} \quad \textrm{for} \quad n \geq 2 \, ,
\ea

\noindent where the $\langle . \rangle$ means, in mathematical terms, the average w.r.t. the noise realizations, and is interpreted in QCD context as the average w.r.t. the target's color fields configurations. The index $c$ denotes cumulants instead of moments of random variables.


\section{Final results}\label{sec:results}

In the result \eqref{derivcumnn} only the constant $C_1 C_2$ remain unknown. We use the same trick as in \cite{Brunet:2005bz} in order to guess it. As explained in \cite{Brunet:2005bz}, the expression of the velocity in the presence of the cut-off \eqref{vco} is still valid if one adds large fluctuations, but with a larger effective front width

\be
F_{eff}(t)=F_0(t)+ \f{3}{\gc} \log F_0(t) \, .
\ee

Then, by matching with \eqref{derivcumnn} the expression obtained, one can find $C_1 C_2$. One gets

\ba \label{derivcum}
\d_t \langle \rho_s(t^2) \rangle &=& v_c - \f{\pi^2 \chi''(\gc)}{2 b \gc v_c F_0^2(t)} + \f{3 \pi^2 \chi''(\gc) \log F_0(t)}{b \gc^2 v_c (F_0(t))^3} \nn
\d_t \langle \rho_s^n(t^2) \rangle_c &=& \f{\ n! \zeta(n) \pi^2 \chi''(\gc)}{\gc^{n+1}  b v_c \ (F_0(t))^3} \quad \textrm{for} \quad n \geq 2  \, .
\ea

Up to now, our results are quite similar to the ones at fixed coupling \cite{Brunet:2005bz}.
However, as the method used in sections \ref{sec:cutoff} and \ref{sec:largef}
is valid only if $F_0(t)\gg 1$, which is equivalent to $\log Y \gg \gc$,
we have only studied the asymptotic behavior of the saturation scale
cumulants, above some very large rapidity $Y_0$. Hence, integrating above $Y_0$,

\ba \label{cumulint} \langle \rho_s(Y) \rangle &=& \langle
\rho_s(Y_0) \rangle + v_c \left(\sqrt{Y}-\sqrt{Y_0}\right) -
\f{\gc \pi^2 \chi''(\gc)}{8 b v_c} \ \int_{\sqrt{Y_0}}^{\sqrt{Y}}
\f{\dd t}{\log^2 t}  \nn & & \qquad \qquad \qquad + \f{3 \gc \pi^2
\chi''(\gc) }{8 b  v_c } \int_{\sqrt{Y_0}}^{\sqrt{Y}} \dd t \
\f{\log\log t }{\log^3 t} \nn \langle \rho_s^n(Y) \rangle_c &=&
\langle \rho_s^n(Y_0) \rangle_c + \f{\ n! \zeta(n) \pi^2
\chi''(\gc)}{8 \gc^{n-2}  b v_c} \ \int_{\sqrt{Y_0}}^{\sqrt{Y}}
\f{\dd t}{\log^3 t} \qquad \textrm{for} \quad n \geq 2  \, . \ea

Two of the integrals in \eqref{cumulint} can be rewritten using the Logarithmic Integral $\textrm{Li}$ \cite{Abra}. Then our final result is

\ba \label{cumul} \langle \rho_s(Y) \rangle &=& \langle
\rho_s(Y_0) \rangle + v_c \left(\sqrt{Y}-\sqrt{Y_0}\right) -
\f{\gc \pi^2 \chi''(\gc)}{8 b v_c} \ \left[\textrm{Li}\left(\sqrt{Y}\right)-\f{\sqrt{Y}}{\log \sqrt{Y}}-\textrm{Li}\left(\sqrt{Y_0}\right)+\f{\sqrt{Y_0}}{\log \sqrt{Y_0}} \right] \nn & & \qquad \qquad \qquad + \f{3 \gc \pi^2
\chi''(\gc) }{8 b  v_c } \int_{\sqrt{Y_0}}^{\sqrt{Y}} \dd t \
\f{\log\log t }{\log^3 t} \nn \langle \rho_s^n(Y) \rangle_c &=&
\langle \rho_s^n(Y_0) \rangle_c + \f{\ n! \zeta(n) \pi^2
\chi''(\gc)}{16 \gc^{n-2}  b v_c} \ \left[\textrm{Li}\left(\sqrt{Y}\right)-\f{\sqrt{Y}}{\log \sqrt{Y}}-\f{\sqrt{Y}}{\left(\log \sqrt{Y}\right)^2}-\textrm{Li}\left(\sqrt{Y_0}\right)+\f{\sqrt{Y_0}}{\log \sqrt{Y_0}}+\f{\sqrt{Y_0}}{\left(\log \sqrt{Y_0}\right)^2} \right]\nn
 & & \qquad \textrm{for} \quad n \geq 2  \, . \ea

We will consider also the following approximated form of our result. Assuming that the logs are not varying much and can be factorized out of the integrals in \eqref{cumulint}, one obtains

\ba \label{cumulapp} \langle \rho_s(Y) \rangle &\simeq& \langle
\rho_s(Y_0) \rangle + v_c \left(\sqrt{Y}-\sqrt{Y_0}\right) -
\f{\gc \pi^2 \chi''(\gc)}{8 b v_c} \
\f{\sqrt{Y}-\sqrt{Y_0}}{\log^2 \sqrt{Y}}  \nn & & \qquad \qquad \qquad + \f{3 \gc \pi^2
\chi''(\gc) }{8 b  v_c } \left(\sqrt{Y}-\sqrt{Y_0}\right)  \
\f{\log\log \sqrt{Y} }{\log^3 \sqrt{Y}} \nn \langle \rho_s^n(Y) \rangle_c &\simeq&
\langle \rho_s^n(Y_0) \rangle_c + \f{\ n! \zeta(n) \pi^2
\chi''(\gc)}{8 \gc^{n-2}  b v_c} \
\f{\sqrt{Y}-\sqrt{Y_0}}{\log^3 \sqrt{Y}} \qquad \textrm{for} \quad n \geq 2  \, . \ea
On Fig. 3, the increase of the variance of the logarithmic saturation scale is plotted for the result \eqref{cumul} and for the approximated result \eqref{cumulapp} as a function of the \emph{effective time} $\sqrt{Y/b}$ associated with the traveling-wave dynamics at running coupling. Precisely, the plotted function is
\be\label{DV}
Var(Y)-Var(Y_0) \equiv  \langle \rho_s^2(Y) \rangle_c - \langle \rho_s^2(Y_0) \rangle_c \, .
\ee
We have also plotted on Fig. 3 the result for another model, mixing fixed and running coupling, in which one takes in \eqref{sBK} $\abar$ at the scale $Q_s(Y)$ and $\alpha_s^2$ for the fluctuation term at the scale $Q_s(Y_0)$. Hence, it features the running coupling mean-field dynamics, but fixed coupling cut-off and fluctuations. It shows a linear rise of the variance of the front as a function of the \emph{effective time}, as expected by analogy with the full fixed coupling case. Running coupling in the noise term strongly reduces the rise of the variance of the front. The truncation \eqref{cumulapp} of our result leads to significantly underestimate that variance, and thus should not be performed.

\begin{figure}\label{fig:disp}
\begin{center}
\includegraphics[width=14cm]{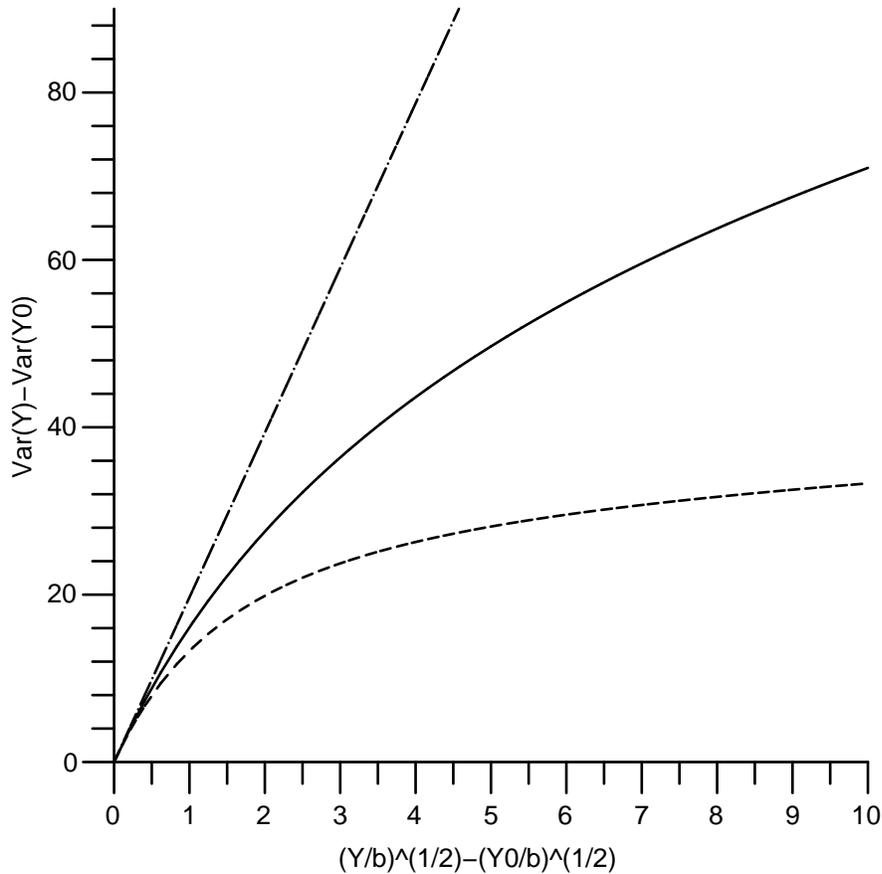}
\caption{Increase of the variance of the logarithmic saturation scale $Var(Y)-Var(Y_0) \equiv  \langle \rho_s^2(Y) \rangle_c - \langle \rho_s^2(Y_0) \rangle_c$, as a function of $\sqrt{Y/b}-\sqrt{Y_0/b}$. Our result \eqref{cumul} is the solid line. The dashed line is the approximation \eqref{cumulapp}, and the dashed-dotted line corresponds to the result for a model with running coupling mean-field dynamics plus fixed coupling noise effects. We have taken here $Y_0=e^{2.5}\simeq 12$.}
\end{center}
\end{figure}

The threshold $Y_0$ has to be adjusted, choosing a compromise between the precision needed and the phenomenological relevance of the rapidity range. We expect that the result \eqref{cumul} catches at least the qualitative features of the full solution for $Y_0\simeq 12$ (which corresponds to consider that $\log Y_0 \simeq 2.5$ is large enough compared to $\gc \simeq 0.63$), and maybe down to $Y_0\simeq 10$. Hence, our result might be relevant for forward physics at the LHC.

In addition, an implicit assumption has been done in our calculations. We have assumed that the wave front is formed down to the low density region where the noise term is relevant, so that it can drive the whole solution. The rapidity interval needed for the solution to form such a wave front is called $Y_{form}$. Hence, we have an additional constraint on $Y_0$ which is $Y_0\ge Y_{form}$. In the present knowledge of QCD evolution equations, no reliable estimate of $Y_{form}$ is available in the running coupling case. It is yet difficult to estimate analytically (even roughly). It requires a much better understanding of the non-asymptotic behavior of the solutions of the BK equation with running coupling\footnote{A rough estimate relying on the asymptotic behavior of the solutions has been attempted in ref.[60], but it does not work in practice.}. Moreover, no numerical simulation of the full BK equation with running coupling and pomeron loops effects has been successful.\\

In our running coupling results \eqref{cumul}, the increase of the cumulants has the same $n$ dependence as in the fixed coupling result of \cite{Brunet:2005bz}. Hence, the calculation of \cite{Marquet:2006xm} is also valid in the running coupling case when the values of the cumulants at $Y_0$ are small compared to the increase between $Y_0$ and $Y$ we have calculated. It happens for $Y\ge Y_0$ if $Y_0 \simeq Y_{form}$ or generically for large enough $Y$. We can conclude that at high rapidity, the dipole-target amplitude is expected to show diffusive scaling
\be\label{diffusivescRC}
{\cal N}(Y, k_T^2) \equiv {\cal N}\left(\tau_d \equiv \f{L-\langle \rho_s(Y) \rangle}{\sqrt{\langle \rho_s^2(Y) \rangle_c}}\right) \, ,
\ee
with $\langle \rho_s(Y) \rangle$ and $\langle \rho_s^2(Y) \rangle_c$ given by \eqref{cumul}.


\section{Conclusion and discussion} \label{sec:conclu}

In this paper we have studied a model extending the BK equation, and expected to include both the main features of fluctuations related to pomeron loops and of running coupling. Using the method developed in \cite{Brunet:2005bz}, we have been able to calculate the high energy asymptotic behavior of its solution, and in particular the asymptotic behavior of the cumulants of the random saturation scale \eqref{cumul}. The \emph{effective time} scale corresponding to the diffusive processes contributing to the front formation or relaxation is very short compared to the \emph{effective time} scale related to the decrease of the coupling. That is why the method of \cite{Brunet:2005bz}, built for fixed coupling, works also in the running coupling case.

Let us compare the results obtained in the running coupling and in the fixed coupling case.  Our result \eqref{derivcum} is rather similar to the one of \cite{Brunet:2005bz} in the fixed coupling case, but some features are modified.  The first difference is that the \emph{effective time} is $\abar Y$ in the fixed coupling case and $\sqrt{Y/b}$ in the running coupling case. That difference is already present in the mean field approximation. The second difference is the logarithmic denominators in the running coupling case, which lower the rise of all the cumulants.

The most subtle difference is in the validity range of the formula \eqref{derivcum}. In our case, it is valid only at high rapidity, whereas in the fixed coupling case, the corresponding formula is valid for any rapidity, provided that the cut-off front is formed, \emph{i.e.} above $Y_{form}$. Hence, in the fixed coupling case, one determines the cumulants completely. This result has been used in \cite{Marquet:2006xm} to demonstrate diffusive scaling \eqref{diffusivescFC}. It comes from a Gaussian approximation, which is valid because the magnitude of the cumulants is not growing too fast with the index of the cumulants.

On the contrary, in the running coupling case, we have been able to calculate only the asymptotic behavior of the cumulants. At smaller rapidity, the front is forming, and this evolution is deterministic. This first regime is characterized by geometric scaling. Then, at $Y=Y_{form}$, when the front is formed up to the region of small $N$ where the fluctuation term becomes relevant, the cumulants start to rise almost linearly, which breaks geometric scaling. Then, as the coupling at the saturation scale decrease, the relevant fluctuations becomes rarer. The rate of growth of the cumulants thus decreases and the evolution of the probability law of the saturation scale slows down, without stopping. Our result \eqref{cumul} concerns only this last regime. \emph{A priori}, a large contribution to the cumulants may come from an intermediate region in rapidity, between $Y_{form}$ and $Y_0$. It might delay the onset of diffusive scaling and lead to a non-scaling regime between the geometric and the diffusive scaling regimes. Hence, our result is not sufficient to predict if diffusive scaling holds in the physical domain in the running coupling case. Nevertheless, for large enough rapidity compared to $Y_0$ the contributions calculated in the present paper will be large compared to the cumulants at $Y_0$, and allow us to predict diffusive scaling \eqref{diffusivescRC} in the high energy limit. Moreover, numerical simulations indicate that the diffusive scaling at fixed coupling is valid not only in the small noise limit, but also at finite noise. One can thus assume that in the running coupling case, the value of the cumulants at $Y_0$ or below should also be compatible with diffusive scaling. Hence our result \eqref{cumul}
\begin{itemize}
\item proves the diffusive scaling (as formulated in \eqref{diffusivescRC}) in the high rapidity limit in the running coupling case;
\item provides a hint that the diffusive scaling should hold for smaller rapidity, perhaps down to $Y_{form}$, in the running coupling case.
\end{itemize}

It is important to note that diffusive scaling can occur only at rapidities larger than $Y_{form}$, whereas geometric scaling is valid at smaller rapidities. Hence, a large $Y_{form}$ could explain the success of the geometric scaling to describe the HERA data. Thus, one infers an experimental lower bound $Y_{form}>9$. If $Y_{form}\leq 12$, it should be possible to observe diffusive scaling \eqref{diffusivescRC} at the LHC. If $Y_{form}$ is larger, geometric scaling will remain valid in the LHC rapidity range. \\

Let us recall and discuss the main assumptions made in the building of our model.
\begin{itemize}
\item The choice of the running coupling at the saturation scale can be \emph{a priori} a matter of debate. It is indeed not completely consistent, as the saturation scale is supposed to be dynamically generated. For that reason, the choice of the running coupling at the saturation scale in the BK equation is the only one giving an equation which can be exactly mapped onto the BK equation with fixed coupling. Except the precise value of the parameters and the fact that the effective time is $\sqrt{Y}$ in the first case and $Y$ in the second, those two equations give the same results. But other choices for the running coupling scale, involving the parent and daughters dipole sizes or momentum scales, give solutions with a slow convergence to the critical asymptotic front and more complicated subasymptotic dynamics. It is a major issue to understand the solution at intermediate rapidity, and hence for phenomenological studies. But it makes no difference for our result, valid only at very high rapidity, when the wave front has been already formed. In the appendix \ref{sec:PDM}, the whole analysis is redone with the choice of the parent dipole momentum as running coupling scale. The results are exactly the same. One can understand this fact in the following way. The relevant region governing the whole solution is the one mentioned as the linear part of the front interior. Indeed, the softer modes are cut-off by the saturation, and the harder by the fluctuations. The size $F_0(t)$ of this region is growing, but only as $\log Y$, whereas this region as a whole moves with the front as $\sqrt{Y}$. Therefore, for two arbitrary logarithmic scales $L$ and $L'$ in that region at large rapidity, $(L-L')/L$ is always suppressed as an inverse power of $Y$. The appendix \ref{sec:PDM} provides an explicit example of this mechanism. Thus we can conclude that any choice for the running coupling scale involving the parent and daughters dipoles sizes or momentum will give the same result. Note that, as mentioned in the introduction, the other NLL contributions cannot be relevant asymptotically \cite{Beuf:2007cw}. Hence, our result \eqref{cumul} can be considered as safe from NLL and higher order effects.
\item Concerning the fluctuations, our approximations are obviously the ones giving the Langevin equation \eqref{sBK}. In particular, the off-diagonal Gaussian noise \cite{Iancu:2005nj} is replaced by a diagonal one, and the dipole-dipole scattering is taken local. These approximations are supposed to provide, at least qualitatively, the same type of fluctuations effects as the full pomeron loops equations. However that is not sure, as neither analytical nor numerical solutions to the full pomeron loops problem are available yet.
\end{itemize}

As an outlook, it would be interesting to obtain analytical results at smaller rapidity than in the present work. However, it would probably require a different framework than the method of \cite{Brunet:2005bz} used here. It would also be sensitive to more details of the model. It would be useful to have a better understanding of the front formation in the running coupling case, in order to have a theoretical prediction for $Y_{form}.$ The main open problem remains however the understanding of the solutions to the full pomeron loops equations.

Note that our results might be relevant outside QCD, as a study of a reaction-diffusion process with space-dependant reaction and diffusion rates. A chemical system with a temperature gradient could show such features.

While completing this work, results of the numerical simulation \cite{Dumitru:2007ew} concerning the extension with running coupling of a 1+1 dimensional reaction-diffusion model \cite{Iancu:2006jw} for high energy QCD have appeared. The main result of \cite{Dumitru:2007ew} is that fluctuations effects are very strongly suppressed in the running coupling case, up to a quite high rapidity. Indeed, $Y_{form}$ is found to be large in that toy model, of the order of a few hundreds, and thus beyond the reach of experiments. It remains to study if such a large $Y_{form}$ is specific to that model, or occurs also in QCD. The results of  \cite{Dumitru:2007ew} and of the present study seem however to be compatible concerning the asymptotic behavior of the cumulants of the saturation scale. Note also that the model used in \cite{Dumitru:2007ew} is a one-dimensional model with similar assumptions concerning the fluctuation term as the one studied here.\\

\begin{acknowledgments}

I would like to thank Robi Peschanski for suggesting to me the subject of this paper, and for many comments and advices. I also thank Edmond Iancu for insightful discussions, in particular about the results of \cite{Dumitru:2007ew} before publication. I also acknowledge Al Mueller and Gregory Soyez for useful remarks.

\end{acknowledgments}

\appendix

\section{Running coupling at the parent dipole momentum scale}\label{sec:PDM}

In this paper, the running coupling has been taken at the saturation scale for convenience. But our results are more general, as they does not depend on the precise scale of the running coupling. As an example, let us present the derivation when one chooses the running coupling scale to be the parent dipole momentum $k_T$.
Inserting
\be
\abar \equiv \f{1}{b L}
\ee
in the equation \eqref{sBK}, one gets, instead of \eqref{sBKr1},
\ba \label{sBKr1bis}
b \ L \ \d_Y {\cal N}(L,Y) &=&   \chi(-\d_L) {\cal N}(L,Y)-{\cal N}^2(L,Y)+\sqrt{\f{\kappa \pi^2 \  {\cal N}(L,Y)}{N_c^2 b^2 \ L^2}} \ \eta(L,Y)  \; , \\
\textrm{with}  & &  \langle \eta(L,Y) \, \eta(L',Y') \rangle = 4 b \  L \ \delta(L \! - \! L') \delta(Y \! - \! Y')\nonumber \; .
\ea
Rewriting \eqref{sBKr1bis} with $t\equiv\sqrt{Y}$ instead of $Y$ and with $\nu(L,t) \equiv \eta(L,t^2) \  \sqrt{t / 2bL}$, one gets
\ba \label{sBKrbis}
\f{b \ L}{2\ t} \ \d_t N(L,t) &=&   \chi(-\d_L) N(L,t)-N^2(L,t)+\sqrt{\f{\bar\kappa \  N(L,t)}{t \ L }} \ \nu(L,t)  \; , \\
\textrm{with}  & &  \langle \nu(L,t) \, \nu(L',t') \rangle = \delta(L \! - \! L') \delta(t \! - \! t') \; ,\nonumber
\ea
in analogy with equation \eqref{sBKr}. The differences between the equations \eqref{sBKr} and \eqref{sBKrbis} are more explicit if we use the comoving coordinate $\xi \equiv L - v(t)t$ to rewrite \eqref{sBKrbis} as
\ba \label{sBKrter}
\f{b}{2} \left(v(t)+\f{\xi}{t}\right) \ \d_t N(L,t) &=&   \chi(-\d_L) N(L,t)-N^2(L,t)+\sqrt{\f{\bar\kappa \  N(L,t)}{t^2 \left(v(t)+\f{\xi}{t}\right)}} \ \nu(L,t)  \; .
\ea
Hence, the cut-off is now for
\be
N(L,t)\sim \f{\bar{\kappa}}{t^2 \left(v(t)+\f{F_0(t)}{t}\right)} \, ,
\ee
instead of $\bar{\kappa}/t^2v(t)$.
In the linear part of the front interior region, we have the equation
\be \label{sBKr11bis} \f{b}{2}\left(v(t)+\f{\xi}{t}\right)[\d_t \bar{N}(\xi, t) -(v(t)+t \dot{v}(t))\d_\xi \bar{N}(\xi,t)]=\chi(-\d_\xi) \bar{N}(\xi,t) \, ,
\ee
instead of \eqref{sBKr11}. In the next step of the method one neglects power suppressed terms. Thus one recovers the equation \eqref{sBKr12}. Hence, the following steps are the same, up to the equations \eqref{contco} and \eqref{derco} in which the explicit value of the cut-off appears. These two equations are replaced by
\ba \label{ContCoBis} \f{A}{\pi } \ \sqrt{\f{\lambda}{\delta v(t)}} \  \sin\left(\pi \sqrt{\f{\delta v(t)}{\lambda}} F_0(t)\right) e^{-\gc F_0(t) } &=& \f{\bar\kappa}{t^2 \ \left(v(t)+\f{F_0(t)}{t}\right)}\\
\label{DerCoBis} \left[-\f{\gc A}{\pi} \  \sqrt{\f{\lambda}{\delta v(t)}} \  \sin\left(\pi \sqrt{\f{\delta v(t)}{\lambda}} F_0(t)\right) + A \cos\left(\pi \sqrt{\f{\delta v(t)}{\lambda}} F_0(t)\right) \right] \  e^{-\gc F_0(t) } &=& -\f{\bar{\g} \bar\kappa}{t^2 \ \left(v(t)+\f{F_0(t)}{t}\right)} \, .
\ea
The $F_0(t)/t$ terms in the denominators of \eqref{ContCoBis} and \eqref{DerCoBis} gives a contribution to $F_0(t)$ only at the order ${\cal O}(F_0(t)/t)$. Hence, the results of the cut-off formalism (\ref{vco}-\ref{F0}) remain the same.

Let us now consider the study of large fluctuations. The equation describing the relaxation after one large fluctuation is again \eqref{sBKr11bis} instead of \eqref{relaxeq}. The $\xi / t$ term is power suppressed in the linear part of the front interior, thus one finds again the equation \eqref{eqtrans}. As the boundary condition for $G$ at the cut-off is not sensitive to the exact value of the cut-off, the following results are the same.

Hence, the results of our asymptotic study are identical if one chooses to take the running coupling either at the saturation scale or at the parent dipole momentum $k_T$. From the comparison of these two cases, one can infer that our results should hold for any reasonable choice of the running coupling scale.



\end{document}